%% file: paper.tex
\documentclass[twoside,twocolumn]{article}

\usepackage{amsmath}
\usepackage{abstract}
\usepackage{titletoc}
\usepackage{titlesec}
\usepackage[normalem]{ulem}

\usepackage{tocloft}
\usepackage[utf8]{inputenc}
\DeclareUnicodeCharacter{2212}{-}
\usepackage{graphicx}
\usepackage{multicol}

\graphicspath{ {figures/} }
\usepackage[backend=biber,style=ieee,sorting=none]{biblatex}
\usepackage[font=footnotesize]{caption}
\usepackage{gensymb}
\usepackage{subcaption}
\usepackage{tabularx}
\usepackage{tabulary}
\usepackage{lscape} 
\usepackage{hhline}
\usepackage[symbol]{footmisc}
\usepackage{booktabs}
\usepackage{multirow}
\usepackage{floatrow}
\usepackage{float}
\usepackage{color}
\usepackage{colortbl}
\usepackage[table,dvipsnames]{xcolor}
\definecolor{Gray}{gray}{0.9}
\definecolor{AB}{rgb}{0.8, 0.9, 1.0}
\usepackage[a4paper, margin=.5In, bottom=1.3In, bindingoffset=0mm]{geometry}
\usepackage{setspace}
\usepackage{txfonts}
\usepackage[nottoc]{tocbibind}

\usepackage{abstract} 

\DeclareUnicodeCharacter{2060}{\nolinebreak}
\usepackage{titling} 

\usepackage{hyperref} 

\pretitle{\begin{center}\Huge\bfseries} 
\posttitle{\end{center}} 
\title{Automated Learning: An Implementation of The A* Search Algorithm over The Random Base Functions} 
\author{%
{Nima Tatari }\\[1ex] 
\normalsize Washington University in St. Louis, Department of Physics \\ 
\normalsize \href{mailto:n.tatari@wustl.edu}{n.tatari@wustl.edu}
}
\date{\today} 
\renewcommand{\maketitlehookd}{%
\begin{abstract}

This letter explains an algorithm for finding a set of base functions. The method aims to capture the leading behavior of the dataset in terms of a few base functions. Implementation of the A-star search will help find these functions, while the gradient descent optimizes the parameters of the functions at each search step. We will show the resulting plots to compare the extrapolation with the unseen data.

\textbf{Keywords:} Automated Learning, Artificial Intelligence, Machine Learning, A-Star Search Algorithm.
\end{abstract}
}

\begin{document}

\maketitle	


\input{sections/1-introduction}
\input{sections/2-methods}

\printbibliography[heading=bibintoc, title={References}]

\end{document}

%% file: sections/1-introduction.tex
\section{Introduction}
The presented idea resembles the asymptotic series representation of a target function as opposed to the convergent series representation, such as the Taylor series. In the asymptotic expansion, the first few terms of the series represent the target function behavior with the desired accuracy in the asymptotic regime \cite{1}. On the other hand, expressing the function by a convergent series would require a large number of terms to get the same accuracy, moreover, the convergent series is a valid representation only inside the disk of convergence on the complex plane. In this method, we reduce the rigor of the analysis and let the algorithm find the leading behavior of the data and its corrections layer by layer. In this sense, the method looks like an asymptotic analysis. However, there exist major differences between the two approaches. We try to develop a link between asymptotic analysis, a search method in artificial intelligence, and pattern recognition in machine learning.

%% file: sections/2-methods.tex
\section{A-star Algorithm}

A-star algorithm, as a search strategy, finds the optimal path between two points on a graph. \cite{2,3} Providing geometric information on the graph helps A-star reduce the number of trials and errors in the search. Such a structure yields the distance between two successive nodes, h(m,n), and the lower limit for the cost between any node n and a goal node f, h(n). As depicted in the graph below, starting from node “i” to find the optimal path to node “f”, as the algorithm explores the node m in the middle, it finds the cost of transitions from m to all its children, h(m,n). In addition, it will calculate h(n) for all the children. We emphasize that if g(m) is the cost spent to reach m, then:

\begin{equation}
\centering
g(m) + h(m, n) + h(n) = g(n) + h(n) = f (n)
\end{equation}

f(n) is the least cost that a path could take to connect i to f while passing through n.

\begin{figure}[ht!]
\centering
\includegraphics[scale=0.3]{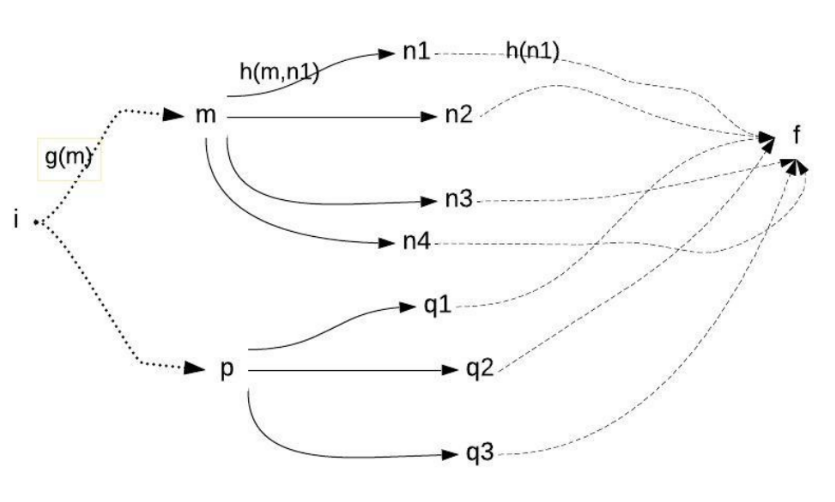}
\caption{Paths to the goal node, h(n): lower bound on the cost from n to f, g(m): cost spent to reach m, h(m,n): exact cost of exploring n from m. Each path from “i” to “f” corresponds to a model.}
\end{figure}

After evaluating f(n) for all the children of m, they are stored in a set called fringe, which contains all the nodes ready for exploration (e.g. n’s and q’s in the diagram above). Then it will choose the node with the least f(n) to explore.

In order to provide a visualization of the algorithm, let us consider a set of nodes, as shown above. We assign a random function with undetermined parameters (e.g. sine, cosine, tanh, etc.) to each node. The algorithm will find the path that connects a subset of these nodes. This path will correspond to a set of base functions to expand the data on that set. While A-star chooses a node among the children, gradient descent optimizes the functions’ parameters at each node exploration.

The algorithm begins by exploring the initial node, which contains the constant function. If the branching is four, there will be four children nodes containing a random function, or product of two or three functions. In the next step, it analyzes each child. Consider the instance where a child is a cosine. The assigned random parameters $\alpha$, $\beta$ in $cos(\alpha x + \beta)$ are subject to optimization by gradient descent so that $c_0 + c_1 cos(\alpha x + \beta)$ produces the best fit to the data points. Note that this function corresponds to a path that connects two nodes, the initial node that contains the constant function, and its child node that contains cosine.

The next step stores the child nodes with optimized parameters in a prioritized order. The error they render while fitting the training and validation(unseen) data determines the priority in exploration.

The subsequent step picks the child node with the least error, among the other nodes stored in the fringe set described earlier.

The presented approach for modeling the data using a few well-known functions considers an ideal path in the graph that would exactly match the data source. Recall that paths in the graph correspond to models, which could describe the data source. Finding the ideal model might remain impossible; nevertheless, a graph endowed with a geometric property, the norm of the difference of two functions as their distance, will navigate A-star towards a perfect model.

\section{Geometry of The Graph}
All the step costs along an optimal path add up to g(n). As a result, finding how to calculate each step cost in the search completes the A-star requirements. The Hilbert space where the paths, which correspond to the functions, live in, equips the graph with geometry because the norm of the difference between any pair of functions serves as the definition for the distance between the corresponding nodes. In addition, the existence of the inner product of the functions validates that the graph contains the elements of a Hilbert space. This leads to the definition of the step cost from the node m to its child n: $h(m,n)= \varphi m(x) − \varphi n(x)$

\begin{figure}[ht!]
\centering
\includegraphics[scale=0.3]{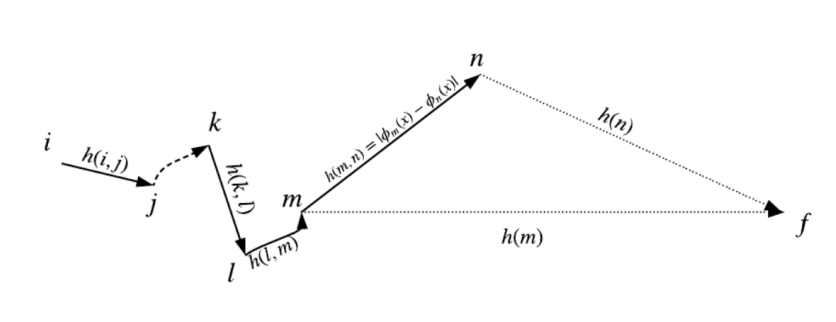}
\caption{The diagram depicts an optimal path from i to n. The heuristics at m and n estimate the cost from these nodes to f, the goal node. The step cost from m to n defines the distance, hence the geometry on the graph. The sum of all the step costs h(p,q) on the path to n adds up to g(n); moreover, the sum of g(n) and h(n) results in f(n) which sets the priority of this path to be chosen or not, for the next exploration at n.}
\end{figure}

\section{Results Without Gradient Descent}
Elimination of the gradient descent at each node expansion requires defining the functions with known, fixed parameters. Since randomness sits at the core of the method, a huge number of functions with random parameters must be generated at each node exploration.

This results in large time complexities. Nevertheless, the first attempts for developing the algorithm utilized this approach, which led to fulfilling results by finding approximately true patterns from the training data set (the red dots in the plots), such as varying oscillation frequencies and amplitudes, as the following diagrams demonstrate. (The term poolSize above the plots shows the number of random functions used in each run, and the term replica means the number of runs to find the paths that each plotted curve shows their weighted sum).

\begin{figure}[ht!]
\centering
\includegraphics[scale=0.4]{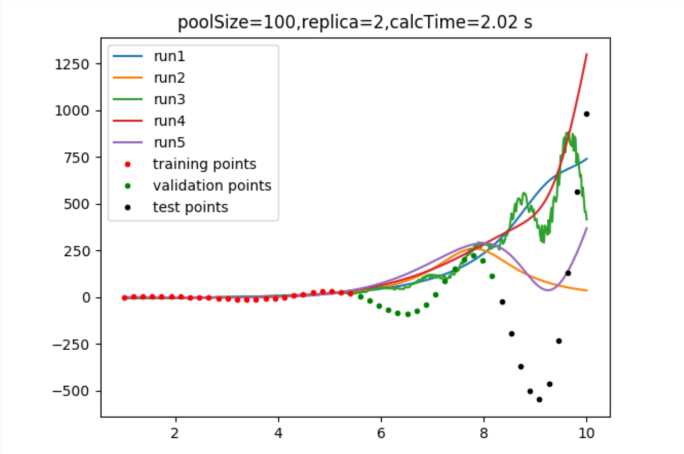}
\caption{Search results with 100 functions in the random set. The models found poorly represent the data trend on the unseen region.}
\end{figure}

\begin{figure}[ht!]
\centering
\includegraphics[scale=0.4]{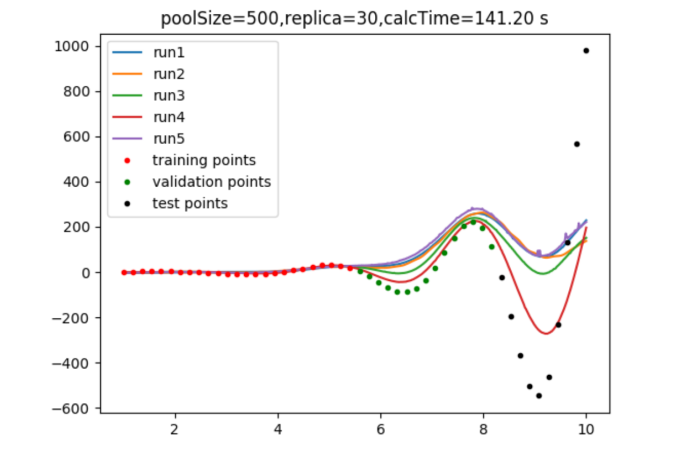}
\caption{The models show better correlations as the random set size increases to 500. However, they fail in predicting the unseen data accurately.}
\end{figure}

\begin{figure}[ht!]
\centering
\includegraphics[scale=0.4]{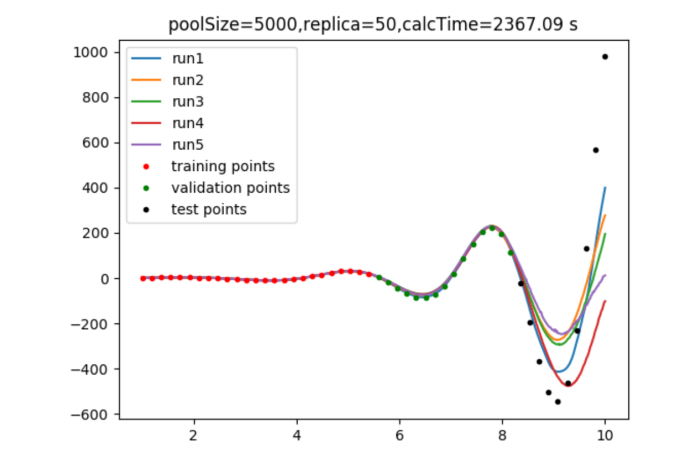}
\caption{Increasing the pool size to 5000.}
\end{figure}

The three diagrams above manifest a shift in the accuracy of predicting the future behavior of the time series, as the number of random functions increases. On the other hand, as this number takes the values of 100, 500, and 5000 respectively, the calculation time for finding five runs takes the trend of 2s, 141s, and 2367s (Recall that each run in these plots corresponds to two replicas, and each replica means a path in the graph. In each run the two paths are weighted by their inverse error on the data).

The next three plots show the results for finding the oscillatory behavior multiplied by polynomial terms $cos(x)(1 + (x − 5) 2 )$. With a pool size of 100 it looks poor in extrapolation, however increasing the size of the random function set to 10000 enables the algorithm to find a model that can predict a part of the unseen data, yet the computation time increases largely.

\begin{figure}[ht!]
\centering
\includegraphics[scale=0.4]{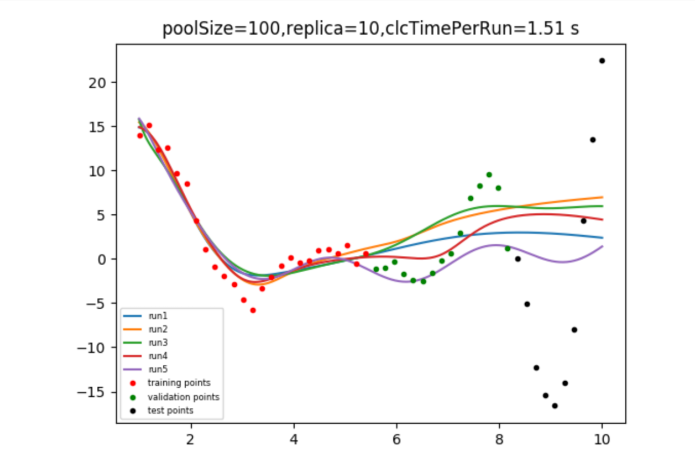}
\caption{Each continuous line shows the result of finding the base functions used to model the data. The models are similar but lack the ability to capture the patterns.}
\end{figure}

\begin{figure}[ht!]
\centering
\includegraphics[scale=0.4]{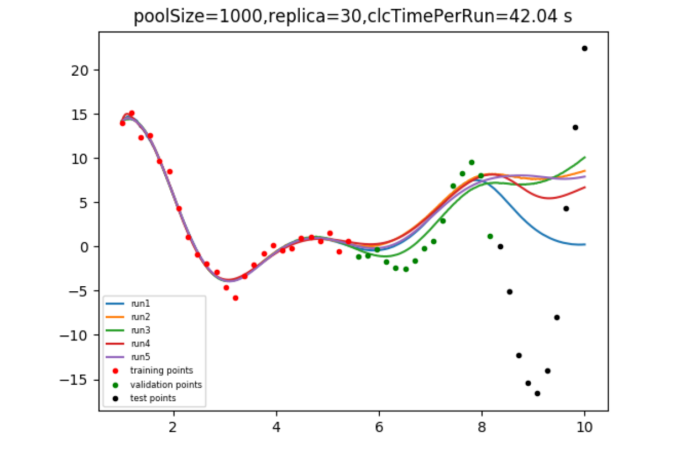}
\caption{Accuracy in prediction improves as the random set size increases.}
\end{figure}

\begin{figure}[ht!]
\centering
\includegraphics[scale=0.4]{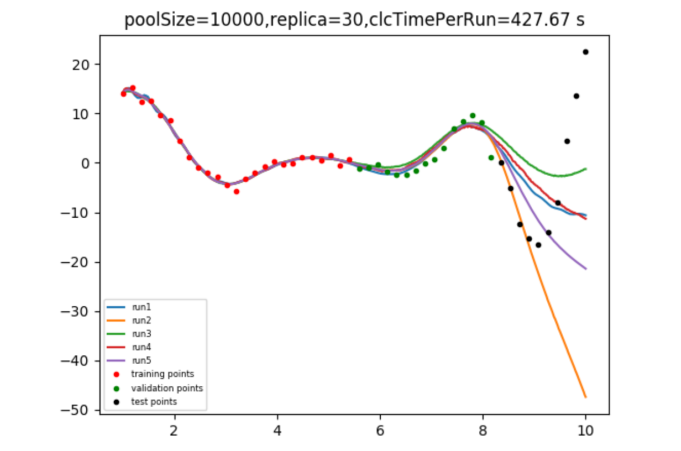}
\caption{Few models are able to extrapolate the data partially, as the random set becomes extremely large.}
\end{figure}

\section{Results Using Gradient Descent}
The examples of the current section correspond to the algorithm armed with gradient descent. The steps of the algorithm follow the procedures described below:

1. Defining the root node in the graph as a constant function, finding its deviation from the training and validation data as $\vec{V}$.

2. Generating a set of random functions with the size of the branching parameter. This set will be used to generate successor nodes.

3. For loop in the random set, analyze each element as a successor of the current node.

4. Given the parent node and a child, optimize the child function(s) parameters using gradient descent. Fitting the resulting base functions to the data, and returning their deviation vectors $\vec{V}$.

5. Exploring the successor with the smallest distance from the goal node $h(n)=|\vec{V}|$.

6. Returning to step 2, if the iteration is smaller than its limit. Otherwise, return the current base set model.

The following figures show the results of the main algorithm. Patterns, such as oscillations, variations in amplitudes, and changes in frequencies are detected by the search scheme.

\begin{figure}[ht!]
\centering
\includegraphics[scale=0.4]{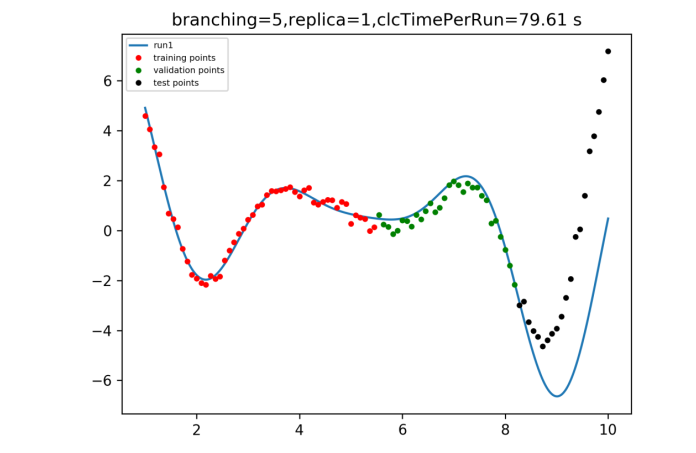}
\caption{Modeling the data behavior after 10 node expansions.}
\end{figure}

\begin{figure}[ht!]
\centering
\includegraphics[scale=0.4]{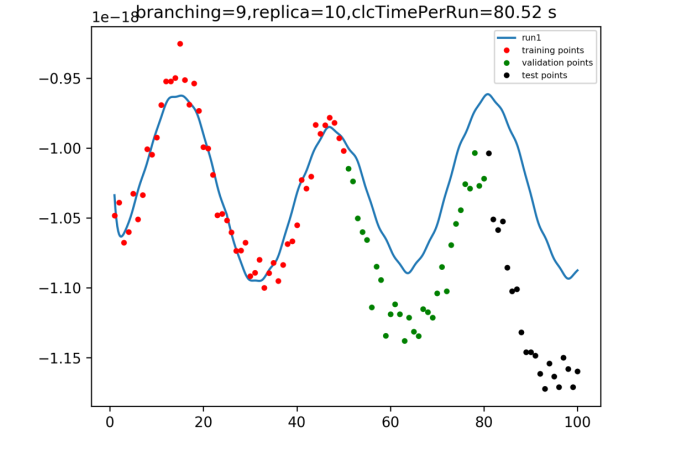}
\caption{Modelling an interval of gravitational wave data detected by LIGO.}
\end{figure}

\section{Time Complexity}
In this section, we evaluate the computation time as a function of different parameters on one of the generated data sets (figure 9). The time complexity with respect to the number of the explored nodes (path length in the graph) with fixed branching number is plotted in figure 11. The plot suggests a polynomial behavior for the time complexity.

\begin{figure}[ht!]
\centering
\includegraphics[scale=0.4]{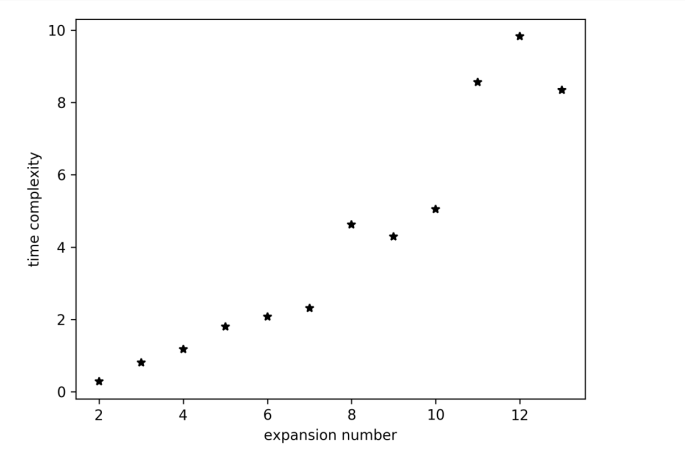}
\caption{Calculation time as the number of explored nodes increases. The number of successors (branching) is set to 8.}
\end{figure}

The next plot investigates the behavior of the complexity in terms of increasing the branching number, which is the number of children of any node, with a fixed number of the total explored nodes.

\begin{figure}[ht!]
\centering
\includegraphics[scale=0.4]{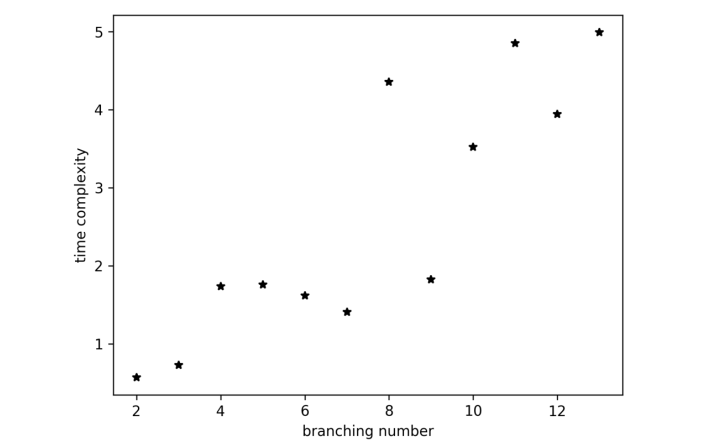}
\caption{Calculation time as the number of successors increases, with the number of totals, explored nodes (path length) set to 6.}
\end{figure}

\section{Further Discussion and Possible Applications}
It remains an important question that how the algorithm applies to real-world problems, in terms of scalability and computational efficiency. It is a true fact that collecting the data sets is the leading factor to train models, regardless of the model complexity. This work forms a platform for future works toward integrating methods from applied mathematics and artificial intelligence and machine learning. For example, in neural networks, one may use A-star to first choose a better set of nonlinear functions. The result can lead to faster pattern recognition, with smaller input data.

\section{Acknowledgement}
The author warmly acknowledges Prof. Zohar Nussinov and Prof. Roman Garnett for their discussions and for providing feedback on this work.